\documentclass[prd,superscriptaddress,amsfonts,amssymb,amsmath,showpacs,onecolumn]{revtex4-2}
\usepackage{bm}
\usepackage{amsfonts}
\usepackage{graphicx}
\usepackage{amsmath}
\usepackage{palatino}
\usepackage{mathpazo}
\usepackage{textcomp}
\usepackage[utf8]{inputenc}
\usepackage[T1]{fontenc}
\usepackage{float}
\usepackage{booktabs}
\usepackage{dcolumn}
\usepackage{multirow}
\usepackage{hyperref}
\hypersetup{colorlinks,citecolor=blue}
\usepackage{amsmath}
\usepackage{xcolor}
\usepackage{orcidlink}
\usepackage[caption=false]{subfig}
\usepackage{commath}
\usepackage{amssymb}
\usepackage{esdiff}
\usepackage{mathtools}
\usepackage{ifthen}
\usepackage{mathtools, amssymb} % !!! Carefull, incompatibility found here, please see below

\usepackage{siunitx}
\usepackage{tikz}
\usetikzlibrary{hobby} % for ..
\usetikzlibrary{arrows.meta} % to control arrow size
\tikzset{>={Latex[length=4,width=4]}} % for LaTeX arrow head
\usetikzlibrary{calc,intersections,decorations.markings}
\usepackage{siunitx}
\usepackage{xcolor} % for colored text

\colorlet{mylightblue}{blue!20}
\colorlet{myblue}{blue!50!black}
\colorlet{mydarkblue}{blue!30!black}
\colorlet{mylightred}{red!10}
\colorlet{myred}{red!50!black}
\colorlet{mydarkred}{red!60!black}
\colorlet{mydarkgreen}{green!30!black}

\tikzset{
  midarr/.style={decoration={markings,mark=at position #1 with {\arrow{stealth}}},postaction={decorate}},
  midarr/.default=0.5
}

\def\jnl@style{\it}
\def\aaref@jnl#1{{\jnl@style#1}}

\def\aaref@jnl#1{{\jnl@style#1}}

\def\aj{\aaref@jnl{AJ}}                   % Astronomical Journal
\def\apj{\aaref@jnl{ApJ}}                 % Astrophysical Journal
\def\apjl{\aaref@jnl{ApJ}}                % Astrophysical Journal, Letters
\def\apjs{\aaref@jnl{ApJS}}               % Astrophysical Journal, Supplement
\def\apss{\aaref@jnl{Ap\&SS}}             % Astrophysics and Space Science
\def\aap{\aaref@jnl{A\&A}}                % Astronomy and Astrophysics
\def\aapr{\aaref@jnl{A\&A~Rev.}}          % Astronomy and Astrophysics Reviews
\def\aaps{\aaref@jnl{A\&AS}}              % Astronomy and Astrophysics, Supplement
\def\mnras{\aaref@jnl{Mon.~Not.~Roy.~Astron.~Soc.}}             % Monthly Notices of the RAS
\def\prd{\aaref@jnl{Phys.~Rev.~D}}        % Physical Review D
\def\prc{\aaref@jnl{Phys.~Rev.~C}}  % Physical Review C
\def\prl{\aaref@jnl{Phys.~Rev.~Lett.}}    % Physical Review Letters
\def\qjras{\aaref@jnl{QJRAS}}             % Quarterly Journal of the RAS
\def\skytel{\aaref@jnl{S\&T}}             % Sky and Telescope
\def\ssr{\aaref@jnl{Space~Sci.~Rev.}}     % Space Science Reviews
\def\zap{\aaref@jnl{ZAp}}                 % Zeitschrift fuer Astrophysik
\def\nat{\aaref@jnl{Nature}}              % Nature
\def\aplett{\aaref@jnl{Astrophys.~Lett.}} % Astrophysics Letters
\def\apspr{\aaref@jnl{Astrophys.~Space~Phys.~Res.}} % Astrophysics Space Physics Research
\def\physrep{\aaref@jnl{Phys.~Rep.}}      % Physics Reports
\def\physscr{\aaref@jnl{Phys.~Scr}}       % Physica Scripta
\def\commat{\aaref@jnl{Comm.~Math.~Phys.}}              % Communications in Mathematical Physics
\def\science{\aaref@jnl{Science}}               % Science
\def\cqg{\aaref@jnl{Classical Quant.~Grav.}}            % Classical and Quantum Gravity
\def\jpcs{\aaref@jnl{JPCS}}                                     % Journal of Physics Conference Series
\def\ijmpd{\aaref@jnl{Int.~J.~Mod.~Phys.~D}}                    % International Journal of Modern Physics D
\def\grg{\aaref@jnl{Gen.~Relat.~Gravit.}}               % General Relativity and Gravitation
\def\rpp{\aaref@jnl{Rep.~Prog.~Phys.}}          % Reports on Progress in Physics
\def\npa{\aaref@jnl{Nucl.~Phys.~A}}        % Nuclear Physics A
\def\lrr{\aaref@jnl{Living Rev.~Rel.}}                   % Living reviews in relativity
\def\jcap{\aaref@jnl{J.~Cosmology Astropart.~Phys.}}    % Journal of cosmology and astroparticle physics
\def\rmp{\aaref@jnl{Rev.~Mod.~Phys.}}   %Reviews of modern physics
\def\epjc{\aaref@jnl{Eur.~Phys.~J.~C}}

\hypersetup{linkcolor=blue}

\allowdisplaybreaks[1]
\begin{document}

\color{black}       %% For one column

\title{Traversable wormholes with charge and non-commutative geometry in the $f(Q)$ gravity}

\author{Oleksii Sokoliuk\orcidlink{0000-0003-4503-7272}}
\email{oleksii.sokoliuk@mao.kiev.ua}
\affiliation{Main Astronomical Observatory of the NAS of Ukraine (MAO NASU),\\
Kyiv, 03143, Ukraine}
\affiliation{Astronomical Observatory, Taras Shevchenko National University of Kyiv, \\
3 Observatorna
St., 04053 Kyiv, Ukraine}
\author{Zinnat Hassan\orcidlink{0000-0002-6608-2075}}
\email{zinnathassan980@gmail.com}
\affiliation{Department of Mathematics, Birla Institute of Technology and
Science-Pilani,\\ Hyderabad Campus, Hyderabad-500078, India.}
\author{P.K. Sahoo\orcidlink{0000-0003-2130-8832}}
\email{pksahoo@hyderabad.bits-pilani.ac.in}
\affiliation{Department of Mathematics, Birla Institute of Technology and
Science-Pilani,\\ Hyderabad Campus, Hyderabad-500078, India.}
\author{Alexander Baransky\orcidlink{0000-0002-9808-1990}}
\email{abransky@ukr.net}
\affiliation{Astronomical Observatory, Taras Shevchenko National University of Kyiv, \\
3 Observatorna
St., 04053 Kyiv, Ukraine}
%%%%%%%%%%%%%%%%%%%%%%%%%%%%%%%%%%%%  DATE  %%%%%%%%%%%%%%%%%%%%%%%%%%%%%%%%%%%%
\date{\today}
\begin{abstract}

We consider modified symmetric teleparallel gravity (STG), in which gravitational Lagrangian is given by the arbitrary function of non-metricity scalar $Q$ to study static and spherically symmetric charged traversable wormhole solutions with non-commutative background geometry. The matter source at the wormhole throat is acknowledged to be anisotropic, and the redshift function has a constant value (thus, our wormhole solution is non-tidal). We study the obtained field equations with the two functional forms of $f(Q)$ STG models, such as linear $f(Q)=\alpha Q+\beta$ and non-linear $f(Q)=Q+mQ^n$ models under Gaussian and Lorentzian distributions. Our analysis found the exact wormhole solutions for the linear STG model only. Also, for the non-linear model, we derived numerically suitable forms of wormhole shape functions directly from the modified Einstein Field Equations (EFEs). Besides, we probed these models via Null, Dominant, and Strong energy conditions with respect to free Modified gravity (MOG)  parameters $\alpha$, $\beta$, $m$, and $n$. We also used Tolman-Oppenheimer-Vokloff (TOV) equation to investigate the stability of wormhole anisotropic matter in considered MOG. Finally, we plot the equation of state. 
\end{abstract}

\maketitle
\section{Introduction}\label{sec:1}
It is well known that wormholes (WHs) are generally the tunnels connecting two widely separated regions in the universe or even two separated universes. Flamm \cite{Flamm} first realized this hypothetical connection in 1916. After that, Einstein and Rosen \cite{Einstein} used his concept and constructed a bridge so-called Einstein-Rosen bridge. Later, in 1957, the term wormhole was introduced by Wheeler, and Misner \cite{Misner}.\\ 
This field of study has been very popular for the last few decades. On the wormholes, there was written a large number of papers, like
\cite{Ellis1973,Bronnikov1973,Morris1988,Hochberg1993,Visser1989,Visser1997,Kim2001,Dadhich2002,Kuhfittig2003}. Among this and other numerous works, one is of special interest - work written by Morris \& Thorne in 1988, which presents humanly traversable spherically symmetric wormholes (in relation to the Einstein-Rosen bridge, which is non-traversable). But, as it turned out, in the conjecture of the Morris-Thorne wormhole, if we consider classical GR gravity, given by the Einstein-Hilbert action below:
\begin{equation}
    \mathcal{S} =\int_\mathcal{M}d^4x\sqrt{-g}\mathcal{R}
\end{equation}
where $\mathcal{R}$ is the Ricci scalar, the so-called Null Energy Condition (NEC) $T_{\mu\nu}k^\mu k^\nu\geq0$ will be violated (here, $T_{\mu\nu}$ is the energy-momentum tensor and $k^\mu$ is a null vector). In GR, Morris-Thorne (MT) wormhole solutions could not be obtained if we consider the non-exotic matter as the matter source. To overcome this issue, researchers used various methods, such as considering the MT wormhole systems where the quantum effect competes with the classical ones \cite{Visser1995LorentzianWF,Gao2016TraversableWV,Maldacena2018,Cceres2019AKV}. Also, we could use additional fields to solve the exotic matter problem \cite{Bronnikov2002,PhysRevD.65.104010,Nicolis2010EnergysAA}. Finally, to overpass the problem of NEC violation, one could assume the modified Einstein-Hilbert action (i.e., modified gravity) because if we modify the EH action, then Einstein Field Equations will differ; thus, the stress-energy tensor will change. Consequently, it could be possible that in one of the viable modified gravity theories, NEC will be satisfied.
\subsection{Modified gravity and wormholes}
General relativity gravity is a good choice at the large scale in our universe because it could sufficiently describe universe evolution. But, as it was noticed during the analysis of recent cosmological observations, GR classical gravity could not describe essential processes, such as cosmological inflation (which occurs at the very early times) or late-time accelerated expansion without additional matter fields such as inflaton. Then, it is beneficial to assume proper EH action modification to describe these processes. For example, one of the most popular choices of the MOG form is $f(\mathcal{R})$ gravity, in which we replace the Ricci scalar with the arbitrary function of the Ricci scalar $f(\mathcal{R})$. Viable $f(\mathcal{R})$ cosmologies coincide very well with the data obtained from the space telescopes, such as \textit{Planck}. For example, Starobinsky model could describe cosmological inflation \cite{Brooker2016,Huang2013,Starobinsky1980} due to presence of squared Ricci scalar in the $f(\mathcal{R})$. Also, with the exponential form of the MOG, we could create the universe with both inflationary and late-time acceleration phases. Finally, even the dark energy problem could be solved \cite{Capozziello2011,Nojiri2017}.\\
In the area of traversable wormholes in modified theories of gravity, many different interesting works have also been studied. In \cite{doi:10.1142/S0217732316501923}, Mazharimousavi and Halilsoy have constructed traversable wormholes in $f(R)$ gravity which is supported by a fluid source and at least satisfies the weak energy conditions. A study on a new class of $f(R)$ gravity model with wormhole solutions and cosmological properties has been presented in \cite{Restuccia2020ANC}. Also, in the same gravity, a note on thin-shell wormholes with charge has been made in \cite{Mazharimousavi2018ANO}. On the extension of $f(R)$ gravity, an interesting work done by Moraes and Sahoo \cite{Moraes:2017mir} on the modeling of wormholes in $f(R,T)$ gravity by considering different relations for their pressure components and different equations of state. Also, the authors of \cite{Mishra2021} discussed wormhole solutions with the quadratic $f(R,T)$ model and studied energy conditions without exotic matter. Further, Sharif and Rani \cite{Sharif/2013} investigated wormhole solutions in $f(T)$ gravity with noncommutative geometry. They observed that the effective energy-momentum tensor is responsible for violating energy conditions to support the nonstandard wormhole. Also, an investigation on traversable wormholes with conformal killing vectors in $f(T)$ gravity have been studied in \cite{Sharif/2014}. For furthermore studies on wormhole geometry, one may check the Refs. \cite{Korolev2020GeneralCO,Mehdizadeh2016NovelTL,Sahoo:2017ual,doi:10.1142/S0218271819501724,Sharif/2013a,Rahaman/2012,Hussain/2021,Sharma1}.\\
The concept of non-commutative geometry is an intrinsic characteristic of the manifold itself, as stated in \cite{Smailagic/2006}, and it can be introduced in GR by modifying the matter source. 
%In \cite{Mathew/2020}, the authors studied in detail the background of both non-commutative distributions. 
It is believed that by using non-commutative geometry, some viewpoints of quantum gravity can be studied mathematically more effectively. An exciting result of the string theory is that the space-time coordinates evolve noncommuting operators on a D-brane \cite{Witten/1995,Seiberg_1999}. Such non-commutative operator are used to encrypted in the commutator $[x^{\mu},x^{\nu}]=i\theta^{\mu\nu}$, where $\theta^{\mu\nu}$ is the the anti-symmetric matrix of dimension $(length)^2$ and it is used to defines the discretization of spacetime \cite{DOPLICHER199439,Smailagic_2004,doi:10.1142/S0217751X09043353}.\\ 
In recent years, non-commutative geometry has become the considerable interest among researchers. It is considered the crucial property of space-time geometry and shows a vital role in different areas. In \cite{Nozari_2009},  Nozaria and Mehdipoura studied ‘Parikh–Wilczek Tunneling from Noncommutative Higher Dimensional Black Holes’ under Lorentzian distribution. Sushkov discussed wormholes supported by phantom energy by employing Gaussian distribution in \cite{PhysRevD.71.043520}. Rahaman et al. \cite{PhysRevD.86.106010} examined wormhole solutions by taking Gaussian distribution in the background and found those wormhole solutions exist in the four as well as in five dimensions only. Moreover, the stability of a particular class of thin-shell wormholes in GR under non-commutative geometry has been studied in \cite{Kuhfittig/2012c}. Also, the BTZ blackhole under non-commutative background has been investigated in \cite{PhysRevD.87.084014}.\\
Our study is focused on recently proposed modified symmetric teleparallel gravity, or so-called $f(Q)$ gravity \cite{Jimenez/2018}. In this kind of MOG, gravitational Lagrangian is described by an arbitrary non-metricity scalar $Q$ function. We focused on this theory because, in recent years, $f(Q)$ MOG gained interest in the community of cosmologists. A large number of works have been studied on this gravity in theoretical and observational directions. We quote, for instance, in \cite{Frusciante/2021,Heisenberg/2020,Bajardi/2020} some cosmological features of $f(Q)$ gravity were investigated, Energy conditions in \cite{Mandal/2020} and also wormhole solutions have been studied in $f(Q)$ gravity in Refs. \cite{Zinnat/2021,Hassan/2021,Mishra/2021}. One may check \cite{Lazkoz/2019,Barrosa/2020,Hassan/2021aa} for more applications of $f(Q)$ gravity.\\
It is worth noting that in the current paper, we investigate the traversable wormhole with noncommutative geometry (both Gaussian and Lorentzian distributions) in the presence of an additional electrostatic field (metric tensor is very similar to the one which describes Reissner-Nordstr\"om charged black hole). \\

\subsection{Article organization}

This article is organized as follows: in the section (\ref{sec:1}), we provide an introduction to the topic of traversable wormholes and, different modified gravity theories, the viability of MOG. In the Section (\ref{sec:2}) we present the formalism of the symmetric teleparallel $f(Q)$ gravity. In the section (\ref{sec:3}), we specify the metric tensor line element of the charged spherically symmetric wormholes and derive modified Einstein Field Equations for such choice of $g_{\mu\nu}$. Furthermore, we also present noncommutative geometry (with both Gaussian and Lorentzian distributions) in this section. In the Section (\ref{sec:4}) we probe the energy conditions of $f(Q)$ charged wormholes with different kinds of noncommutative geometries and different forms of $f(Q)$ function.  Additionally, in the section (\ref{sec:5}), we show how the equation of state parameter $\omega$ changes with the change of radial coordinate $r$, in the section (\ref{sec:6}), we derive the fair values of MOG parameters, for which wormhole is stable. Finally, in the last section (\ref{sec:7}), we provide the concluding remarks about the key topics of our investigation.

\section{Formalism of the \texorpdfstring{$f(Q)$}{} Gravity} \label{sec:2}

In the $f(Q)$ gravity, the total Einstein Hilbert action is given:
\begin{equation}
    \mathcal{S}[g_{\mu\nu},\Gamma,\Psi] = \mathcal{S}_g+\mathcal{S}_m =  \frac{1}{2\kappa}\int d^4x \sqrt{-g} \bigg[f(Q)+2\kappa\mathcal{L}_m[g_{\mu\nu},\Psi_i]\bigg]
\end{equation}
where $f(Q)$ is arbitrary function of non-metricity scalar $Q$, $\kappa$ is gravitational constant, further we will assume that $\kappa=1$, and finally $\mathcal{L}_m[g_{\mu\nu},\Psi_i]$ is the Lagrangian density of all perfect fluid (or spinor, gauge boson) matter fields $\Psi_i$ coupled to gravity $g_{\mu\nu}$. Finally, in the action integral above $\Gamma$ is the well-known affine connection, but in the case of modified STG this connection is not metric compatible, torsion free. Firstly, we obviously want to define non-metricity tensor \cite{Jimenez/2018}
\begin{equation}
    Q_{\alpha\mu\nu} = \nabla_\alpha g_{\mu\nu}
\end{equation}
where $\nabla_{\alpha}$ is covariant derivative and $g_{\mu\nu}$ is charged wormhole metric tensor, which we will define in the next section. Fundamental quantity for MOG of our consideration is non-metricity scalar
\begin{equation}
Q = -Q_{\alpha \mu \nu}P^{\alpha \mu \nu}
\end{equation}
The non-metricity conjugate is \cite{Khyllep2021}
\begin{equation}
P^\alpha\;_{\mu\nu}=\frac{1}{4}\left[-Q^\alpha\;_{\mu\nu}+2Q_{(\mu}\;^\alpha\;_{\nu)}+Q^\alpha g_{\mu\nu}-\tilde{Q}^\alpha g_{\mu\nu}-\delta^\alpha_{(\mu}Q_{\nu)}\right],
\end{equation}
where $Q_{\alpha}=Q_{\alpha}\;^{\mu}\;_{\mu}$ and $\tilde{Q}_\alpha=Q^\mu\;_{\alpha\mu}$ are traces of non-metricity tensor.\\
Then, while we already defined all of the necessary, we could proceed to the derivation of the Einstein Field Equations by varying the EH action integral w.r.t. metric tensor $g_{\mu\nu}$: 
\begin{equation}
\frac{2}{\sqrt{-g}}\nabla_\gamma\left(\sqrt{-g}\,f_Q\,P^\gamma\;_{\mu\nu}\right)+\frac{1}{2}g_{\mu\nu}f \\
+f_Q\left(P_{\mu\gamma i}\,Q_\nu\;^{\gamma i}-2\,Q_{\gamma i \mu}\,P^{\gamma i}\;_\nu\right)=-T_{\mu\nu},
\label{1c}
\end{equation}
where $f_Q\equiv\frac{df}{dQ}$. Also, by varying the action w.r.t. the affine connection $\Gamma^\alpha_{\,\,\,\,\mu\nu}$ we obtain:
\begin{equation}
\nabla_\mu \nabla_\nu \left(\sqrt{-g}\,f_Q\,P^\gamma\;_{\mu\nu}\right)=0.
\end{equation}
%\begin{equation}
%    \nabla_\mu\nabla_\nu (\sqrt{-g}f_QP^{\mu\nu}_{}_{\alpha})=0
%\end{equation}
Therefore, we could go ahead and present the traversable wormhole spacetime in the next section.

\section{Traversable wormholes in \texorpdfstring{$f(Q)$}{} gravity with non-commutative geometry}\label{sec:3}

Firstly, as usual we want to present the spherically symmetric, static traversable wormhole spacetime (in the spherical coordinates) preserving a charge $\mathcal{Q}$ \cite{Morris1988,Kim2001ExactSO}: 
\begin{equation}
\begin{gathered}
ds^2 = -\left(1+\frac{\mathcal{Q}^2}{r^2}\right) dt^2+\left(1-\frac{b}{r}+\frac{\mathcal{Q}^2}{r^2}\right)^{-1}dr^2+r^2d\theta^2 + r^2\sin^2\theta d\phi^2,
\label{eq:1}
\end{gathered}
\end{equation}
where $b(r)$ is the shape function that determines the shape of the wormhole. Shape function defines the geometry of the traversable wormhole, and must obey following (in)equalities: i) $b-r=0$ at the WH throat ($r=r_0$), ii) $\frac{b-rb'}{b^2}>0$, iii) $b'<1$, iiii) $\lim_{r\to\infty}\frac{b}{r}=0$ (because of the asymptotically flat background). Also, because we want to obtain only horizonless and non-singular solutions, $e^{2\Omega(r)}$ (here $\Omega(r)$ is the redshift function) must always be finite, and also, from the \cite{Morris1988}, tidal forces of the wormholes must be bearably small. Because of that conditions, we could consider the Zero Tidal Forces (ZTF) kind of traversable wormhole. It is worth noticing that the line element (\ref{eq:1}) connects Morris-Thorne SS spacetime and Reissner-Nordstr\"om spacetime, so if $\mathcal{Q}=0$, we will have Morris-Thorne wormhole without charge, and if $b=0$, we will have Reissner-Nordstr\"om black hole (because of the fact that $r>0$, we don't have the singularity in the charged WH spacetime).\\
The stress tensor for an anisotropic fluid compatible with
spherical symmetry is
\begin{equation}
T_{\mu}^{\nu}=(\rho+P_t)u_{\mu}u^{\nu}+P_t \delta_{\mu}^{\nu}+(P_r-P_t)v_{\mu}v^{\nu},
\label{1a}
\end{equation}
where, $\rho$ denotes the energy density. $u_{\mu}$ and $v_{\mu}$ are the four velocity vector and unitary space-like vectors, respectively. Also both are satisfy the conditions $-u_{\mu}u^{\nu}=v_{\mu}v^{\nu}=1$. $P_r$ and $P_t$ denotes the radial and tangential pressures and both are functions of radial coordinate $r$.\\
For the metric \eqref{eq:1}, the non-metricity scalar $Q$ can be written as
\begin{equation}
    Q=\frac{(\mathcal{Q}^2-r b)}{r^3}\left(\frac{r \left(b'-2\right)+b}{r (r-b)+\mathcal{Q}^2}-\frac{2 \mathcal{Q}^2}{\mathcal{Q}^2+r^2}+\frac{2}{r}\right).
    \label{1b}
\end{equation}
Now, by using Eqs. \eqref{eq:1},\eqref{1a} and \eqref{1b} in Eq. \eqref{1c}, we get non zero components of the field equations are
\begin{multline}
\label{3a}
\rho=\frac{\mathcal{Q}^2+r^2-r\,b}{2\,r^4}\left[\frac{f\,r^4}{\mathcal{Q}^2+r^2-r\,b}+2\,r\,f_{QQ}Q^{'}\left(\frac{r^2}{\mathcal{Q}^2+r^2-r\,b}-1\right)+\right.\\ \left. f_Q\left(\left(2-\frac{2\mathcal{Q}^2\,r}{\mathcal{Q}^2+r^2} \right) \left( \frac{r^2}{\mathcal{Q}^2+r^2-r\,b}-1 \right)+\frac{(\mathcal{Q}^2+r^2-r\,b)(2\mathcal{Q}^2-r\,b+r^2\,b^{'})}{(\mathcal{Q}^2+r^2-r\,b)^2} \right)  \right],
\end{multline}
\begin{multline}
\label{3b}
P_r=-\frac{\mathcal{Q}^2+r^2-r\,b}{2\,r^4}\left[\frac{f\,r^4}{\mathcal{Q}^2+r^2-r\,b}+2\,r\,f_{QQ}Q^{'}\left(\frac{r^2}{\mathcal{Q}^2+r^2-r\,b}-1\right)+\right.\\ \left. f_Q\left(\frac{4\,\mathcal{Q}^2\,r}{\mathcal{Q}^2+r^2}+\left( \frac{r^2}{\mathcal{Q}^2+r^2-r\,b}-1 \right) \left(2-\frac{2\mathcal{Q}^2\,r}{\mathcal{Q}^2+r^2}+\frac{2\mathcal{Q}^2-r\,b+r^2\,b^{'}}{\mathcal{Q}^2+r^2-r\,b} \right) \right) \right],
\end{multline}
\begin{multline}
\label{3c}
P_t=-\frac{\mathcal{Q}^2+r^2-r\,b}{4\,r^3}\left[\frac{2\,f\,r^3}{\mathcal{Q}^2+r^2-r\,b}-2\,r\,f_{QQ}Q^{'}+\right.\\ \left. f_Q\left(-\frac{4\mathcal{Q}^2\,r\,(\mathcal{Q}^2+2\,r)}{(\mathcal{Q}^2+r^2)^2}-\frac{4\mathcal{Q}^2\,(2\,r\,b-r^2-2\,\mathcal{Q}^2)}{(\mathcal{Q}^2+r^2)(\mathcal{Q}^2+r^2-r\,b)}+\frac{2\,\mathcal{Q}^2-r\,b+r^2\,b^{'}}{r(\mathcal{Q}^2+r^2-r\,b)}\left(\frac{2\,r^2}{\mathcal{Q}^2+r^2-r\,b}-\frac{2\,\mathcal{Q}^2\,r}{\mathcal{Q}^2+r^2} \right) \right) \right].
\end{multline}
One can verify the above field equations will reduce to Einstein's GR when $f(Q)=Q$ and charge $\mathcal{Q}=0$.\\
Finally, we will proceed to the charged wormhole non-commutative geometry behavior.

\subsection{Non-commutative geometry}

In the current article, to simplify the calculations, we will be using the so-called non-commutative geometry ansatz. Usually, non-commutative geometry is used in GR for the replacement of point-like structures as the smeared object (which allows us to eliminate the divergencies). This smearing effect could be achieved by the replacement of the Gaussian distributions of minimal length $\sqrt{\theta}$ with the Dirac delta function. In \cite{Mathew/2020}, Schneide and DeBenedictis deeply examined the background of both non-commutative distributions. In the next sections, we shall discuss the physical analysis of wormhole solutions under non-commutative Gaussian and Lorentzian distributions. For this purpose, we consider the Gaussian and Lorentzian distributions of the energy densities for the point-like gravitational source are given below \cite{Barros2020,NICOLINI2006547}:
\begin{equation}
    \rho(r) = \frac{Me^{-\frac{r^2}{4\theta}}}{8\pi^{3/2}\theta^{3/2}} 
    \label{eq:3.3}
\end{equation}
%And also there is present so-called Lorentzian distributions:
\begin{equation}
    \rho(r) = \frac{\sqrt{\theta}M}{\pi^2(\theta+r^2)^2}
    \label{eq:3.12}
\end{equation}
where $\theta$ is the non-commutativity parameter. $M$ is the smearing mass distribution, and it could be a diffused centralized
an object such as a wormhole \cite{Ponce/2003}.
% Further we will probe both Gaussian and Lorentzian distributions.

\section{Constraining charged WH's from energy conditions}\label{sec:4}

\subsection{Energy Conditions}

We will probe the following energy conditions in the current paper:
\begin{itemize}
    \item Null Energy Condition (NEC): $\rho+p_r \geq 0\land \rho+p_t \geq 0$
    \item Weak Energy Condition (WEC): $\rho\geq0$ and $\rho+p_r \geq 0 \land \rho+p_t \geq 0$
    \item Strong Energy Condition (SEC): $\rho + p_r + 2p_t \geq 0$
    \item Dominant Energy Condition (DEC): $\rho \geq |p_r| \land \rho \geq |p_t|$
\end{itemize}
As we know, in the GR, if traversable wormholes exist, there always must present so-called exotic matter at the throat, which violates Null Energy Condition (minimal requirement of WEC and SEC). In this paper, we will investigate the energy conditions of the wormhole in the viable $f({Q})$ cosmologies in the presence of non-commutative geometry. 

\subsection{Gaussian distribution}

In this section, we are going to probe the different energy conditions for our charged traversable wormhole with various $f(Q)$ models and with Gaussian distribution energy density.
\subsubsection{Linear model \texorpdfstring{$f(Q) = \alpha Q + \beta$}{}}
\begin{figure}[!htbp]
    \centering
    \includegraphics[scale=0.7]{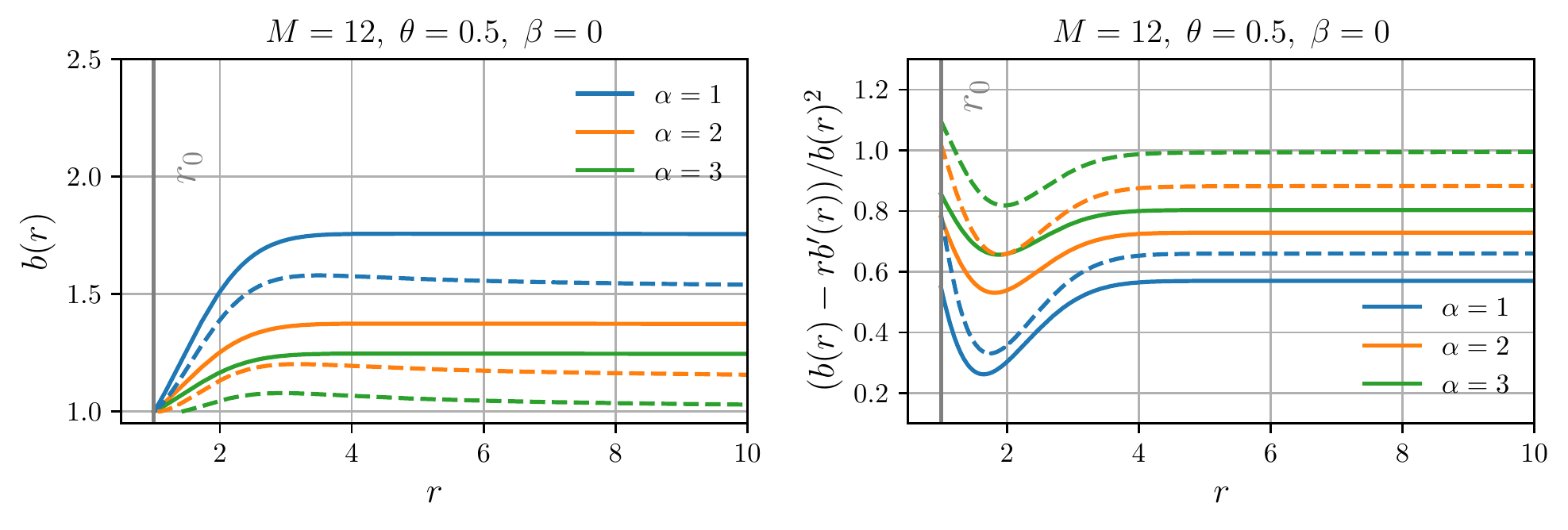}
    \caption{Charged wormhole shape function and flaring-out condition for the linear $f(Q)$ gravity (Gaussian distribution) with varying $\alpha$ and vanishing $\beta$. In order to obtain the solutions, we assumed that $M=12$, $\theta=0.5$, moreover on the plots solid line represents the solution with $\mathcal{Q}=0.1$ and dashed with $\mathcal{Q}=0.5$}
    \label{fig:1}
\end{figure}

\begin{figure}[!htbp]
    \centering
    \includegraphics[scale=0.7]{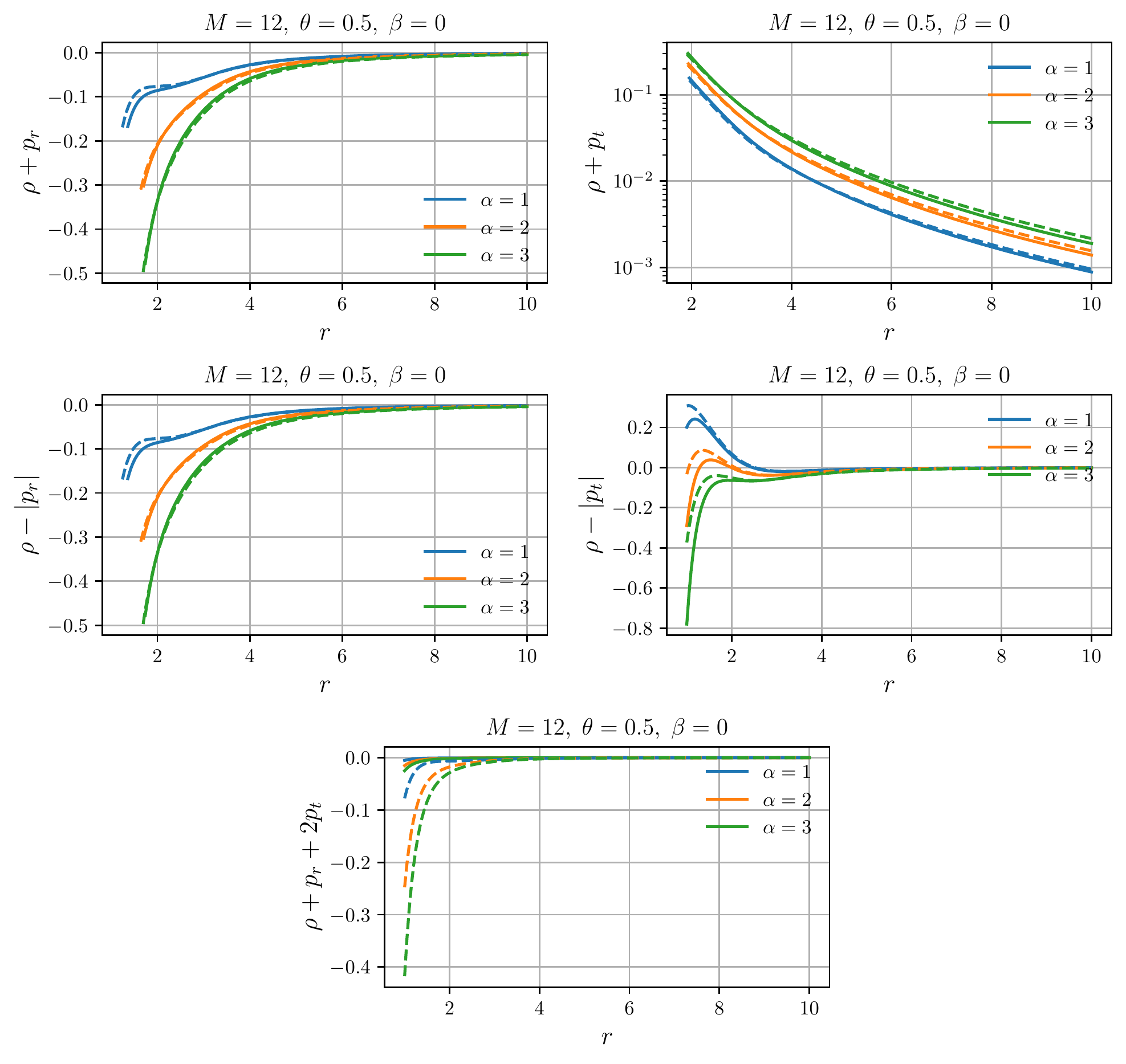}
    \caption{Linear $f(Q)$ gravity Null, Dominant, and Strong energy conditions for the charged traversable wormhole spacetime. For simplicity, we assumed that $M=12$ and $\theta=0.5$, and also we took $\beta=0$}
    \label{fig:2}
\end{figure}

As for the first model of STG, we consider following simplest linear form of $f(Q)$ function \cite{Mustafa2022}:
\begin{equation}
    f(Q) = \alpha Q + \beta,
\end{equation}
where $\alpha$ and $\beta$ are free parameters. It is known that the linear functional form of $f(Q)$ retrieves the symmetric teleparallel equivalent to general relativity, which enables us to compare our solutions to their fundamental level.\\
Under this specific $f(Q)$ model, we compare Eqs. \eqref{3a} and \eqref{eq:3.3} and able to get the differential equation for the shape function $b(r)$ given by
\begin{equation}
    %\label{4A}
    b^{'}(r)=\frac{M\,r^2\,e^{-\frac{r^2}{4 \theta }}}{\alpha\,\left(4\pi \theta\right)^\frac{3}{2}}-\left(\frac{\mathcal{Q}}{r}\right)^2-\frac{\beta\,r^2}{2\,\alpha},
\end{equation}
and it's solution is given by
\begin{equation}
\label{4a}
    b(r)=\frac{1}{12 \alpha  r}\bigg(\frac{3 M r\, \text{Erf}\left(\frac{r}{2 \sqrt{\theta }}\right)}{\pi }-\frac{3 M r^2 e^{-\frac{r^2}{4 \theta }}}{\pi ^{3/2} \sqrt{\theta }}+12 \alpha  \mathcal{Q}^2-2 \beta  r^4\bigg)+ c_1,
\end{equation}
where $c_1$ is the integrating constant.\\
To extract $c_1$, we impose the throat condition $b(r_0)=r_0$ in Eq. \eqref{4a} and obtain
\begin{equation}
\label{4B}
    c_1=r_0-\bigg(\frac{3 M \,r_0\, \text{Erf}\left(\frac{r_0}{2 \sqrt{\theta
   }}\right)}{\pi }-\frac{3 M r_0^2 e^{-\frac{r_0^2}{4 \theta }}}{\pi ^{3/2}
   \sqrt{\theta }}+12 \alpha  \mathcal{Q}^2-2 \beta  r_0^4\bigg)(12 \alpha  r_0)^{-1}.
\end{equation}
Inserting Eq. \eqref{4B} into Eq. \eqref{4a}, we get the final version of $b(r)$ under Gaussian distribution given below
\begin{equation}
    \label{4C}
    b(r)=\mathcal{Q}^2 \left(\frac{1}{r}-\frac{1}{r_0}\right)+\frac{M \left(\text{Erf}\left(\frac{r}{2 \sqrt{\theta }}\right)-\text{Erf}\left(\frac{r_0}{2 \sqrt{\theta }}\right)\right)}{4 \pi  \alpha }+\frac{\frac{3 M r^2 e^{-\frac{r_0^2}{4 \theta }}}{\pi ^{3/2} \sqrt{\theta }}-\frac{3 M r r_0 e^{-\frac{r^2}{4 \theta }}}{\pi ^{3/2} \sqrt{\theta }}-2 \beta  r^3 r_0+12 \alpha  r_0^2+2 \beta\,r_0^4}{12 \alpha  r_0}.
\end{equation}
 We plotted the graphs for shape function and flaring out condition on the Figure (\ref{fig:1}) with varying $\alpha$, $\mathcal{Q}$ and vanishing $\beta$, $M=12$ and $\theta=0.5$. As one could notice, our solution for the Gaussian distribution is physically viable since the flaring out condition is satisfied everywhere within our Lorentzian manifold.\\
Inserting Eq. \eqref{4C} into Eqs. \eqref{3a}-\eqref{3c}, we get the following components
\begin{equation}
    \label{4b}
    \rho=\frac{Me^{-\frac{r^2}{4\theta}}}{8\pi^{3/2}\vartheta^{3/2}},
\end{equation}
\begin{multline}
\label{4c}
p_r=\bigg(-4 \left(3 \alpha  c_1 \left(\mathcal{Q}^2 (1-2 r)+r^2\right)+\mathcal{Q}^2 r^2 (6 \alpha +\beta  r (r+1))+\beta  r^5\right)-\frac{3 M \left(\mathcal{Q}^2 (1-2 r)+r^2\right)
   \text{Erf}\left(\frac{r}{2 \sqrt{\theta }}\right)}{\pi }\\
  +\frac{3 M r \left(\mathcal{Q}^2 (1-2 r)+r^2\right) e^{-\frac{r^2}{4 \theta }}}{\pi ^{3/2} \sqrt{\theta }}\bigg)\bigg/\bigg(12 r^3
   \left(\mathcal{Q}^2+r^2\right)\bigg),
\end{multline}
\begin{multline}
\label{4d}
p_t=\bigg(8 \bigg(3 \alpha  c_1 \left(\mathcal{Q}^4 \left(-2 r^2+r+1\right)+\mathcal{Q}^2 (2-3 r) r^2+r^4\right)+\mathcal{Q}^4 r^2 (6 \alpha  (r-1)+\beta  r ((r-2) r-2))\\
+2 \mathcal{Q}^2 r^4 (3 \alpha -2 \beta  r)-2
   \beta  r^7\bigg)+6 M \left(\mathcal{Q}^4 \left(-2 r^2+r+1\right)+\mathcal{Q}^2 (2-3 r) r^2+r^4\right) \text{Erf}\left(\frac{r}{2 \sqrt{\theta }}\right)\pi^{-1}-3 M r
   e^{-\frac{r^2}{4 \theta }} \\ \times\left(-\mathcal{Q}^4 (r-1) \left(2 \theta +r^2+4 \theta  r\right)-\mathcal{Q}^2 r^2 (r (6 \theta +(r-2) r)-4 \theta )+r^6+2 \theta  r^4\right)\pi ^{-3/2} \theta
   ^{-3/2}\bigg)\bigg/\bigg(48 r^3 \left(\mathcal{Q}^2+r^2\right)^2\bigg),
\end{multline}
where $c_1$ is defined in Eq. \eqref{4B}.\\
In addition, we probed the Null, Dominant, and Strong energy conditions in the Figure (\ref{fig:2}). Unfortunately, because of the non-commutative geometry, NEC is violated for radial pressure but validated for the tangential one. DEC is violated at each point of spacetime for radial pressure as well as the SEC. However, tangential DEC could be obeyed for relatively small and positive values of $\alpha$.

\subsubsection{Non-linear model \texorpdfstring{$f(Q) = Q + mQ^n$}{}}

\begin{figure}[!htbp]
    \centering
    \includegraphics[scale=0.7]{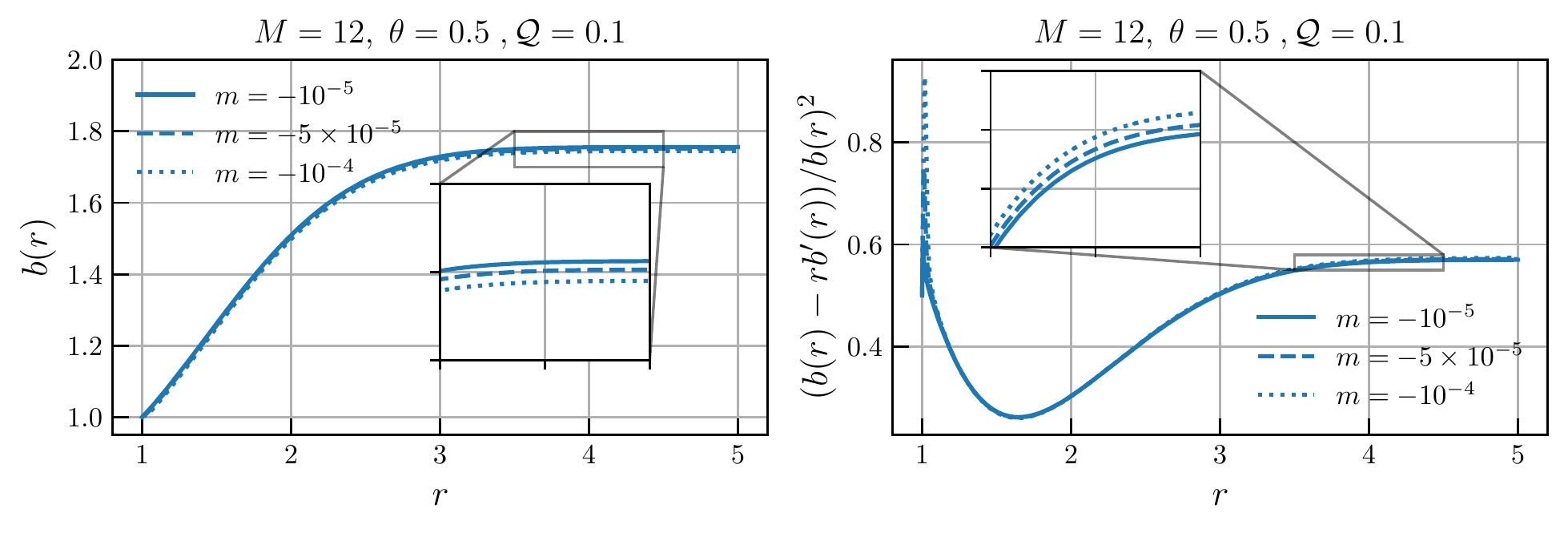}
    \caption{Shape function and flaring-out conditions for non-linear $f(Q)$ model with Gaussian smeared mass distribution and $M=12$, $\theta=0.5$, $\mathcal{Q}=0.1$ and $r_0=1$}
    \label{fig:333}
\end{figure}

\begin{figure}[!htbp]
    \centering
    \includegraphics[scale=0.7]{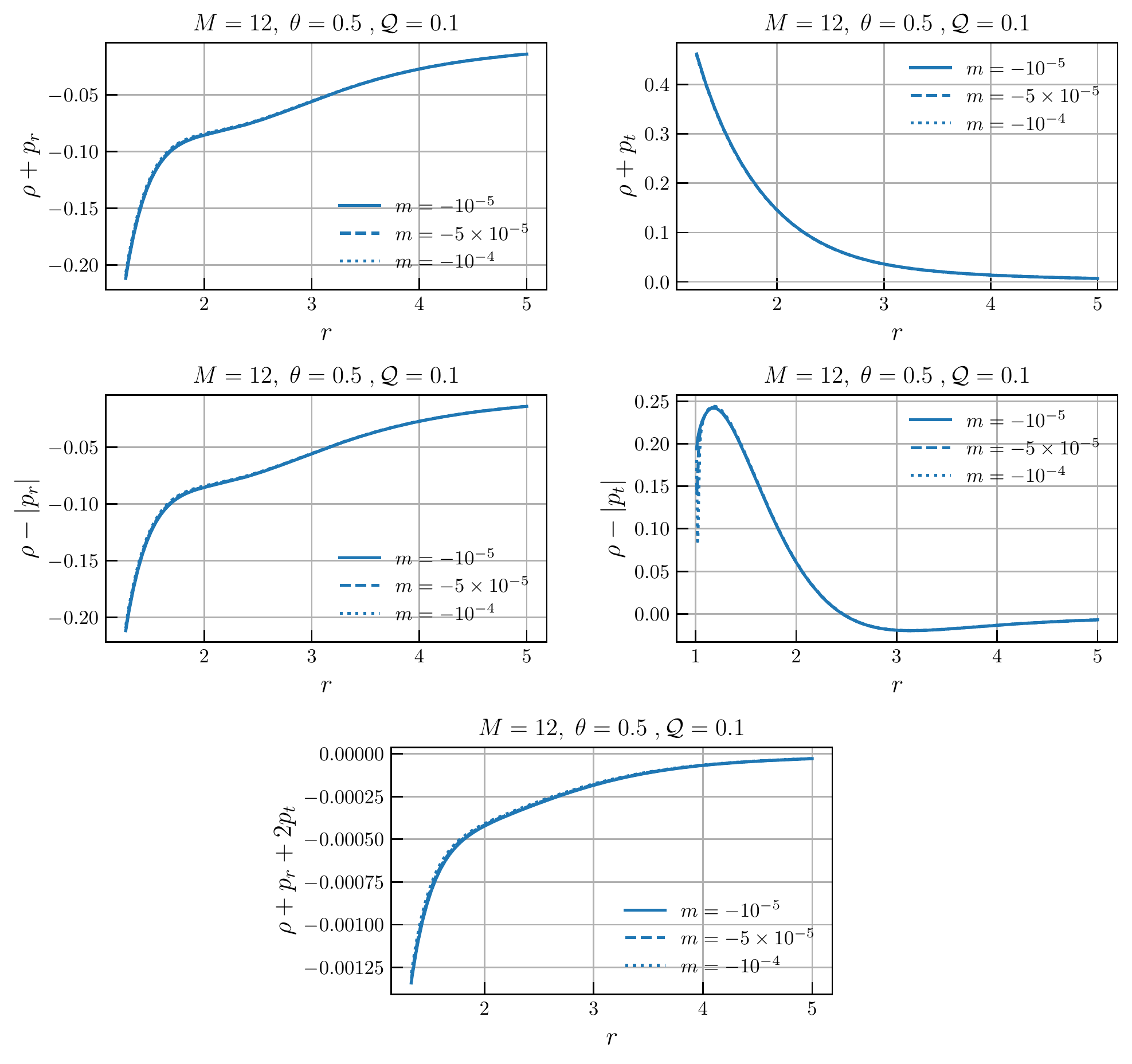}
    \caption{Non-linear $f(Q)$ gravity Null, Dominant, and Strong energy conditions for the charged traversable wormhole spacetime (with the Gaussian distribution of energy density). For simplicity, we assumed that $M=12$ and $\theta=0.5$, and also we took $\mathcal{Q}=0.1$}
    \label{fig:444}
\end{figure}

Throughout this subsection, we have used non-linear form of the MOG function $f(Q)$ \cite{Shekh:2021ule}:
\begin{equation}
    f(Q) = Q+mQ^n,
\end{equation}
where $m$ and $n$ are free parameters. A particular form of this model has been used in different studies. Lin and Zhai \cite{Lin:2021uqa} have studied stellar structure with polytropic Equation of state (EoS) by considering $n=2$ and found that $m<0$ provides support to more stellar masses while positive $m$ reduces the amount of matter of the star. Also,  Banerjee $et\,al.$ investigated wormhole solutions in \cite{Banerjee2021} with this non-linear model and concluded that wormhole solutions could not exist for this specific functional form. They also fixed $n=2$. Motivated by the above, we continue our study with the non-linear quadratic form of the $f(Q)$ model.\\
Due to the high complexity of the field equations, we could not find the exact charged wormhole solutions analytically with this specific model. Hence, we are bounded to fix some initial conditions to study the CWH solutions. We consider the initial conditions so that these conditions satisfy all the necessary conditions of shape functions. The considered initial conditions are given by
\begin{equation}
\label{4b2a}
    b(r_0)=r_0\,\,\,\,\, \text{and}\,\,\,\,\, b'(r_0)=1/2,
\end{equation}
where $r_0$ is the throat radius.\\
We numerically solved the equations \eqref{3a} and \eqref{eq:3.3} with initial conditions \eqref{4b2a} and studied the behavior of shape functions and energy conditions. We illustrated shape function and flaring-out condition solutions with constant WH charge $\mathcal{Q}=0.1$ on the Figure (\ref{fig:333}). As we have noticed during the numerical analysis, the flaring-out condition for the shape function in the non-linear $f(Q)$ gravity is satisfied at the CWH throat, and for small values of charge, $\mathcal{Q}$ is also satisfied at the asymptotically flat CWH region.
As well, we have solved the various energy conditions and plotted numerical solutions in the Figure (\ref{fig:444}) respectively. As it turned out, unfortunately, for every positive value of WH charge $\mathcal{Q}$, NEC is violated for the radial pressure and validated for the tangential one. The DEC situation was the same, and SEC was violated near the WH throat.

\subsection{Lorentzian distribution}

This subsection will probe the different energy conditions for our charged traversable wormhole with various $f(Q)$ models and with Lorentzian distribution energy density.

\subsubsection{Linear model \texorpdfstring{$f(Q) = \alpha Q + \beta$}{}}
In this case, we equate the Eq. \eqref{3a} and \eqref{eq:3.12}, and obtain the differential equation for $b(r)$ as follows
\begin{equation}
    \label{4A}
    b^{'}(r)=\frac{M\,r^2\,\sqrt{\theta}}{\alpha\,\pi^2\,(\theta+r^2)^2}-\left(\frac{\mathcal{Q}}{r}\right)^2-\frac{\beta\,r^2}{2\,\alpha},
\end{equation}
after solving, we are able to find the equation for shape function $b(r)$ given by
\begin{equation}
    \label{4e}
    b(r)=c_2-\frac{\sqrt{\theta } M r}{2 \pi ^2 \alpha  \left(\theta +r^2\right)}-\frac{M \tan ^{-1}\left(\frac{\sqrt{\theta }}{r}\right)}{2 \pi ^2 \alpha }+\frac{\mathcal{Q}^2}{r}-\frac{\beta  r^3}{6 \alpha },
\end{equation}
where $c_2$ is the integrating constant. As usual, we could derive it with the help of throat condition $b(r_0)=r_0$ in Eq. \eqref{4e}
\begin{equation}
\label{4f}
    c_2 = \frac{\sqrt{\theta } M r_0}{2 \pi ^2 \alpha  \left(\theta +r_0^2\right)}+\frac{M \tan ^{-1}\left(\frac{\sqrt{\theta
   }}{r_0}\right)}{2 \pi ^2 \alpha }-\frac{\mathcal{Q}^2}{r_0}+\frac{\beta  r_0^3}{6 \alpha }+r_0.
\end{equation}
Substituting Eq. \eqref{4f} into Eq. \eqref{4e}, we obtain $b(r)$ given by
\begin{equation}
    \label{4g}
    b(r)=\frac{1}{6} \Bigg(6 \mathcal{Q}^2 \left(\frac{1}{r}-\frac{1}{r_0}\right)+\frac{-\frac{3 \sqrt{\theta } M r}{\pi ^2 \left(\theta +r^2\right)}+\frac{3 \sqrt{\theta } M r_0}{\pi ^2 \left(\theta +r_0^2\right)}-\beta  r^3+\beta  \text{r1}^3+6 \alpha  r_0}{\alpha }+\frac{3 M \left(\cot ^{-1}\left(\frac{r_0}{\sqrt{\theta }}\right)-\cot ^{-1}\left(\frac{r}{\sqrt{\theta }}\right)\right)}{\pi ^2 \alpha }\Bigg).
\end{equation}
\begin{figure}[!htbp]
    \centering
    \includegraphics[scale=0.7]{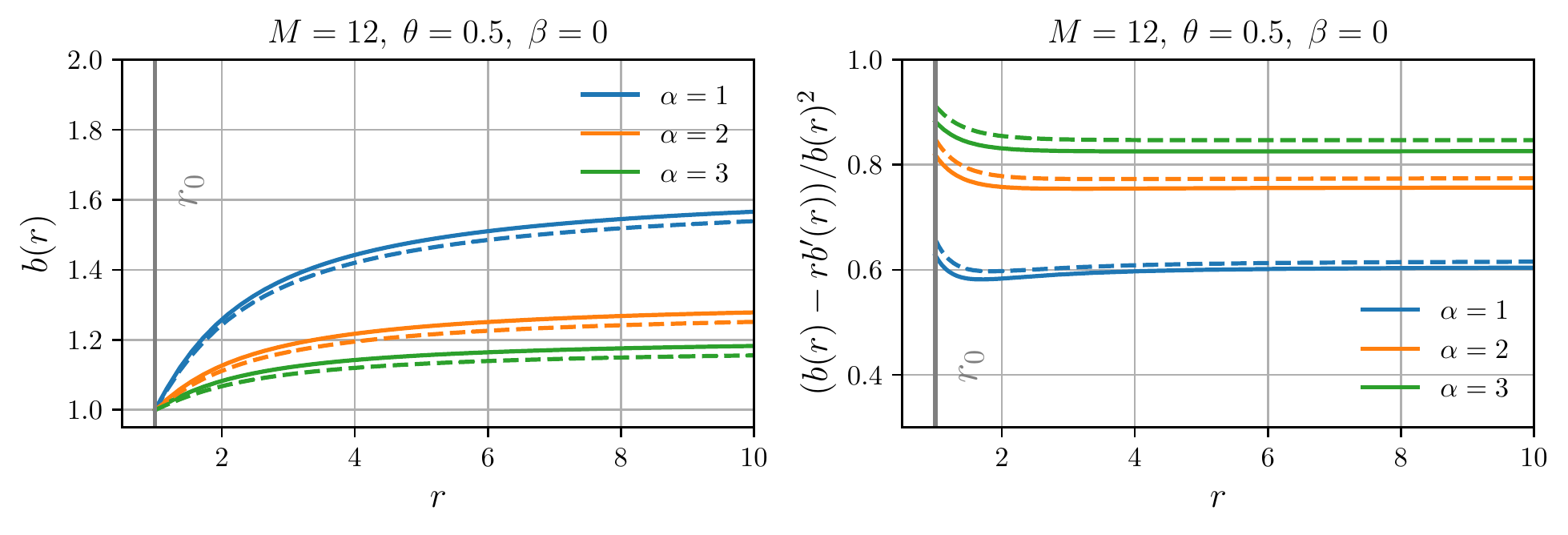}
    \caption{Charged wormhole shape function and flaring-out condition for the linear $f(Q)$ gravity (Lorentzian distribution) with varying $\alpha$ and vanishing $\beta$. In order to obtain the numerical solutions, we assumed that $M=12$, $\theta=0.5$, moreover on the plots solid line represents the solution with $\mathcal{Q}=0.1$ and dashed with $\mathcal{Q}=0.2$}
    \label{fig:3}
\end{figure}
In Fig. \ref{fig:3}, we have depicted the behavior of shape functions and flaring out conditions under asymptotic background. One can find that shape function showing positively increasing behavior and flaring out condition is also satisfied as $b^{'}(r_0)<1$ at $r=r_0$. Here we fix some parameters $M=12$, $\theta=0.5$, $\beta=0$ and $r_0=1$ with varying $\alpha$ and $\mathcal{Q}$.\\
Again, substituting Eq. \eqref{4g} into the Eqs. \eqref{3a}-\eqref{3c}, we obtain the components of energy-momentum tensor are
\begin{equation}
    \label{4h}
    \rho=\frac{\sqrt{\theta } M}{\pi ^2 \left(\theta +r^2\right)^2},
\end{equation}
\begin{multline}
\label{4i}
p_r=\bigg(-6 \pi ^2 \alpha  c_2 \left(\mathcal{Q}^2 (1-2 r)+r^2\right) \left(\theta +r^2\right)+3 M \left(\mathcal{Q}^2 (1-2 r)+r^2\right) \left(\theta +r^2\right) \cot
   ^{-1}\left(\frac{r}{\sqrt{\theta }}\right)\\
   +r \left(3 \sqrt{\theta } M \left(\mathcal{Q}^2 (1-2 r)+r^2\right)-2 \pi ^2 r \left(\theta +r^2\right) \left(\mathcal{Q}^2 (6 \alpha +\beta  r
   (r+1))+\beta  r^3\right)\right)\bigg)\bigg/\bigg(6 \pi ^2 r^3 \left(\mathcal{Q}^2+r^2\right) \left(\theta +r^2\right)\bigg),
\end{multline}
\begin{multline}
    \label{4j}
    p_t=-\bigg(6 \pi ^2 \alpha  c_2 \left(\mathcal{Q}^4 (r-1) (2 r+1)+\mathcal{Q}^2 r^2 (3 r-2)-r^4\right) \left(\theta +r^2\right)^2-3 M \left(\mathcal{Q}^4 (r-1) (2 r+1)+\mathcal{Q}^2 r^2 (3 r-2)\right.\\ \left.
    -r^4\right)\left(\theta +r^2\right)^2 \cot ^{-1}\left(\frac{r}{\sqrt{\theta }}\right)+r \bigg(3 \sqrt{\theta } M \left(\mathcal{Q}^4 (-(r-1)) (\theta +r (2 \theta +r (2 r+3)))+\mathcal{Q}^2 r^2 (2 \theta
   +r ((6-5 r) r-3 \theta )) \right.\\ \left.
   +r^4 \left(\theta +3 r^2\right)\right)+2 \pi ^2 r \left(\theta +r^2\right)^2 \left(\mathcal{Q}^4 (\beta  r (2-(r-2) r)-6 \alpha  (r-1))+2 \mathcal{Q}^2 r^2 (2 \beta 
   r-3 \alpha )+2 \beta  r^5\right)\bigg)\bigg)\\
   \bigg/\bigg(12 \pi ^2 r^3 \left(\mathcal{Q}^2+r^2\right)^2 \left(\theta +r^2\right)^2\bigg),
\end{multline}
where $c_2$ is defined in Eq. \eqref{4f}.
\begin{figure}[!htbp]
    \centering
    \includegraphics[scale=0.7]{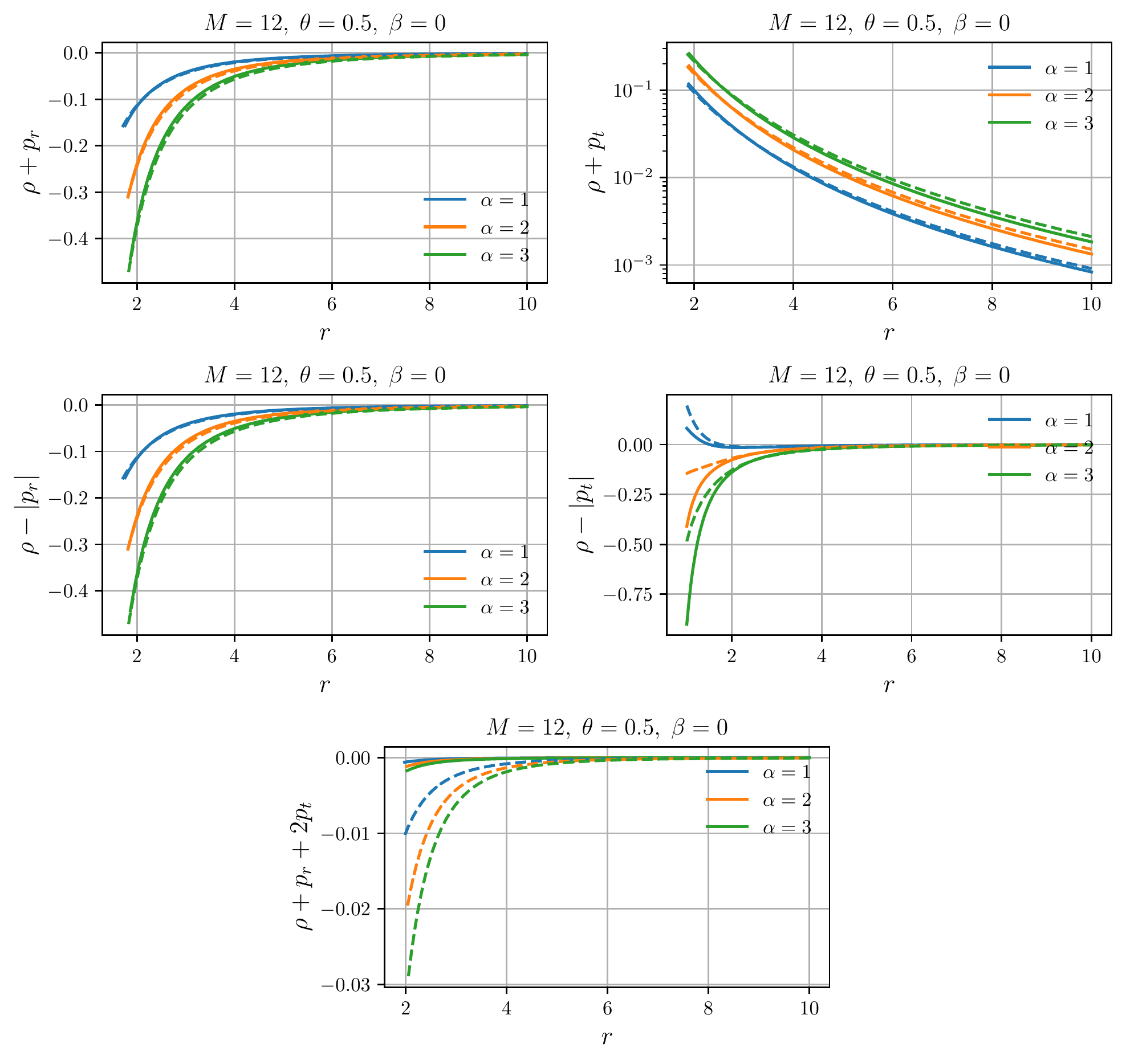}
    \caption{Linear $f(Q)$ gravity Null, Dominant, and Strong energy conditions for the charged traversable wormhole spacetime (with the Lorentzian distribution of energy density). For simplicity, we assumed that $M=12$ and $\theta=0.5$, and also we took $\beta=0$}
    \label{fig:4}
\end{figure}
Furthermore, we analytically constrained CWH spacetime by the Null, Dominant, and Strong energy conditions in the Figure (\ref{fig:4}). As it was revealed, in relation to the Gaussian distribution, in the Lorentzian one with the linear $f(Q)$ model NEC was violated for the radial and validated for tangential pressure with any $\mathcal{Q}\geq0$. Dominant Energy Condition, in turn, was violated for both pressure kinds if $\alpha\gg0$. Finally, SEC was also violated even further from the charged wormhole throat (because of the smeared WH mass).

\subsubsection{Non-linear model \texorpdfstring{$f(Q) = Q + mQ^n$}{}}

\begin{figure}[!htbp]
    \centering
    \includegraphics[scale=0.7]{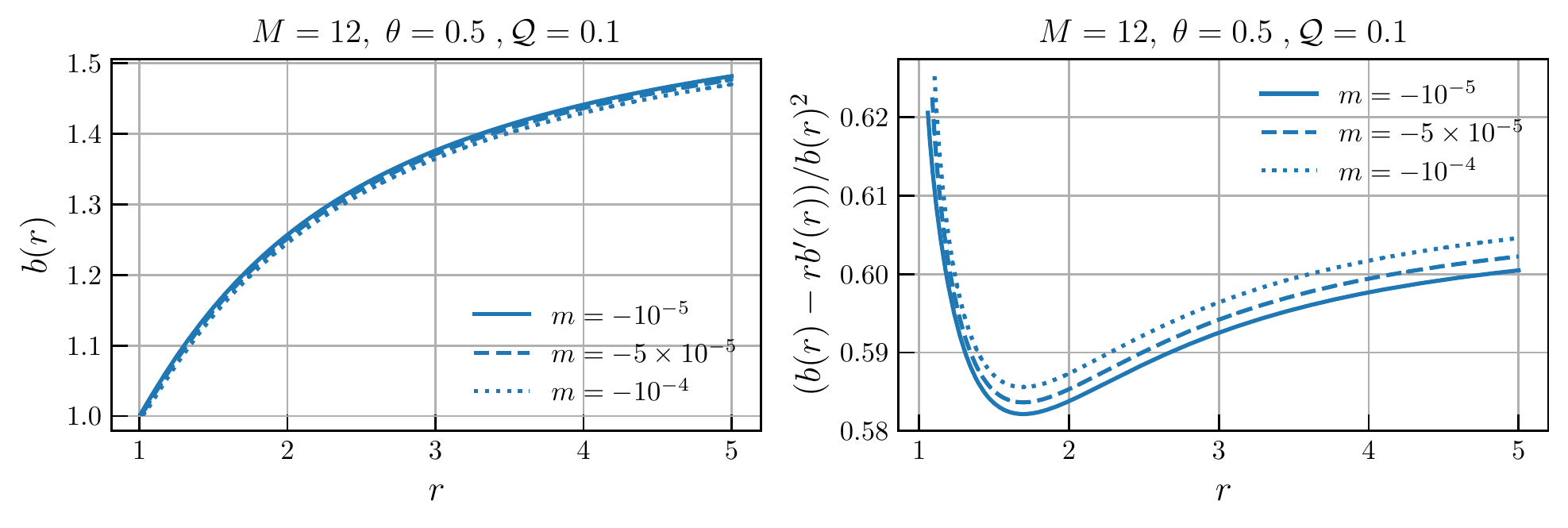}
    \caption{Shape function and flaring-out conditions for non-linear $f(Q)$ model with Lorentzian smeared mass distribution and $M=12$, $\theta=0.5$, $\mathcal{Q}=0.1$ and $r_0=1$}
    \label{fig:777}
\end{figure}

\begin{figure}[!htbp]
    \centering
    \includegraphics[scale=0.7]{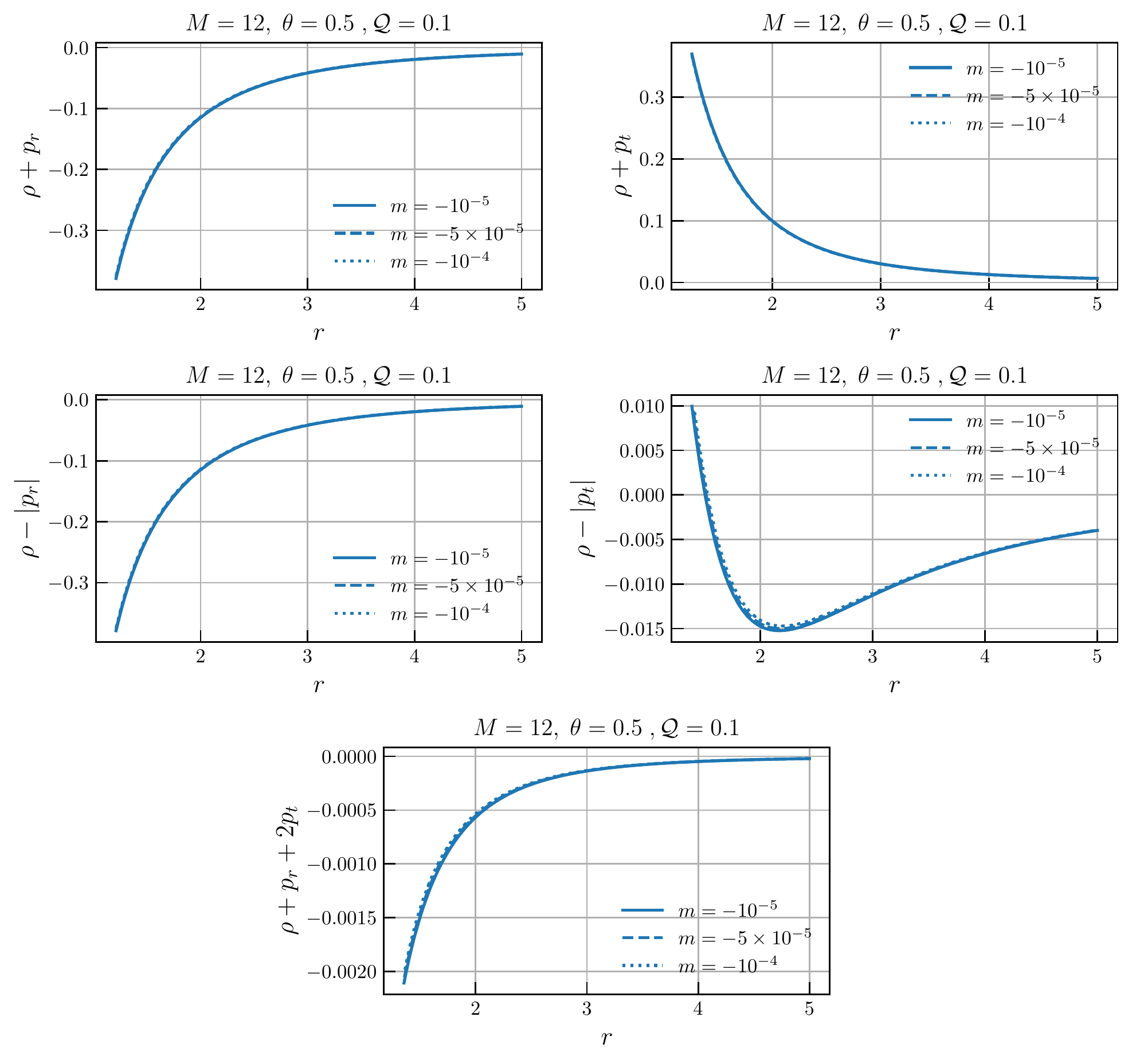}
    \caption{Non-linear $f(Q)$ gravity Null, Dominant, and Strong energy conditions for the charged traversable wormhole spacetime (with the Lorentzian distribution of energy density). For simplicity, we assumed that $M=12$ and $\theta=0.5$, and also we took $\mathcal{Q}=0.1$}
    \label{fig:888}
\end{figure}

Our final model is the non-linear $f(Q) = Q + mQ^n$ MOG with the Lorentzian distribution. Here, we also adopted the same initial conditions used in the Gaussian distribution. Consequently, in the Figure (\ref{fig:777}), we present the charged traversable wormhole shape function and its flaring-out condition. It is necessary to notice that the flaring-out condition for the CWH spacetime with the Lorentzian non-commutative geometry is validated both near the throat and at the aymptotically flat region for any value of $\mathcal{Q}\geq0$.
As it was unveiled, for the non-linear $f(Q)$ charged wormhole with the Lorentzian distribution energy density, all energy conditions except tangential NEC (NEC for radial, DEC for both radial and tangential pressures, and SEC) were violated for any $\mathcal{Q}\geq0$ (for more details and numerical representation, see Figure (\ref{fig:888})).

\section{Equation of State}\label{sec:5}

\begin{figure}[!htbp]
    \centering
    \includegraphics[scale=0.7]{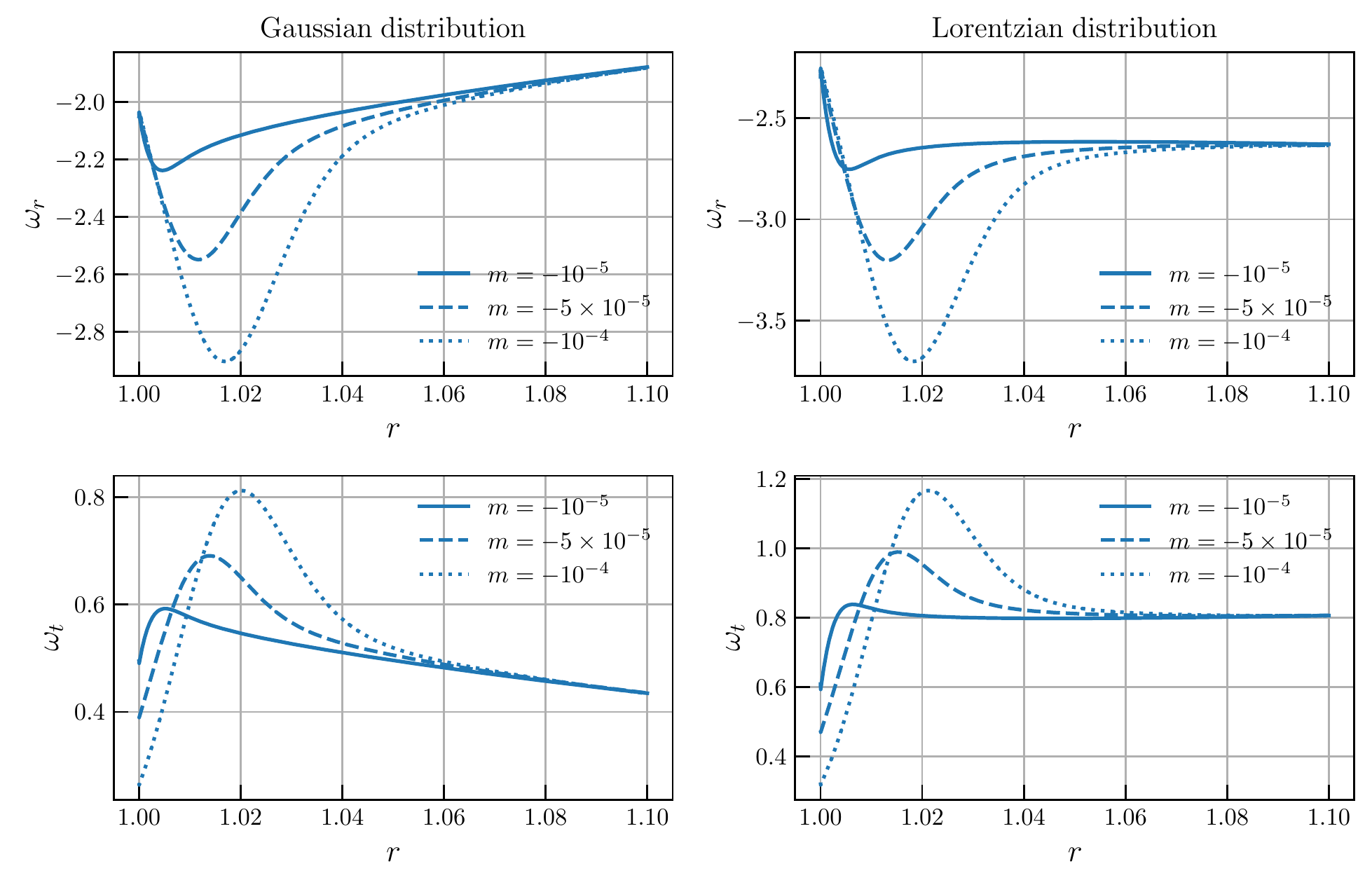}
    \caption{Non-linear $f(Q)$ gravity equation of state parameter for the charged traversable wormhole spacetime with both Gaussian and Lorentzian smeared mass distribution. In order to obtain results graphically, we assumed that $M=12$ and $\theta=0.5$, and also we took $\mathcal{Q}=0.1$}
    \label{fig:1000}
\end{figure}

In this section, we shall discuss the behavior of EoS parameter $\omega$ with different charges $Q$. For this, we have considered the relation between energy density $\rho$ and radial, tangential pressures $p_r$, $p_t$ as follows
\begin{equation}
\label{p}
p_r=\omega_r\rho,\quad p_t=\omega_t\rho,
\end{equation}
where $\omega_r(r)$ and $\omega_t(r)$ are the radial and tangential EoS parameters, respectively.\\
For linear model, $\omega_r(r)$ and $\omega_t(r)$ can be obtained under Gaussian distribution from Eqs. (\ref{4b}-\ref{4d}) along with the relation (\ref{p}) define by
\begin{multline}
    \label{p1}
    \omega_r(r)=\frac{1}{3 M r^3 \left(\mathcal{Q}^2+r^2\right)}\left(-8 \pi ^{3/2} \theta ^{3/2} e^{\frac{r^2}{4 \theta }} \left(3 \alpha c_1 \left(\mathcal{Q}^2 (1-2 r)+r^2\right)+\mathcal{Q}^2 r^2 (6 \alpha +\beta  r (r+1))+\beta  r^5\right)\right.\\ \left.
    -6 \sqrt{\pi } \theta ^{3/2} M \left(\mathcal{Q}^2 (1-2 r)+r^2\right) e^{\frac{r^2}{4 \theta }} \text{Erf}\left(\frac{r}{2 \sqrt{\theta }}\right)+6 \theta  M r \left(\mathcal{Q}^2 (1-2 r)+r^2\right)\right),
\end{multline}
\begin{multline}
   \label{p2}
   \omega_t(r)=\frac{1}{6 M r^3 \left(\mathcal{Q}^2+r^2\right)^2}\left(8 \pi ^{3/2} \theta ^{3/2} e^{\frac{r^2}{4 \theta }} \left(3 \alpha\,c_1 \left(\mathcal{Q}^4 \left(-2 r^2+r+1\right)+\mathcal{Q}^2 r^2 (2-3 r)+r^4\right)+2 \mathcal{Q}^2 r^4 (3 \alpha -2 \beta  r)\right.\right.\\ \left.\left.
   +\mathcal{Q}^4 r^2 (6 \alpha  (r-1)+\beta  r ((r-2) r-2))-2 \beta  r^7\right)+6 \sqrt{\pi } \theta ^{3/2} M \left(\mathcal{Q}^4 \left(-2 r^2+r+1\right)+\mathcal{Q}^2 r^2 (2-3 r)+r^4\right) e^{\frac{r^2}{4 \theta }}\right.\\ \left. 
   \text{Erf}\left(\frac{r}{2 \sqrt{\theta }}\right)-3 M r \left(-\mathcal{Q}^2 r^2 (r (6 \theta +(r-2) r)-4 \theta )-\mathcal{Q}^4 (r-1) \left(2 \theta +r^2+4 \theta  r\right)+2 \theta  r^4+r^6\right) \right),
\end{multline}
respectively.\\
Similarly under Lorentzian distribution, $\omega_r(r)$ and $\omega_t(r)$ can be obtained from Eqs. (\ref{4h}-\ref{4j}) with the relation (\ref{p}) given by
\begin{multline}
   \label{p3}
   \omega_r(t)=-\frac{1}{6 \sqrt{\theta } M r^3 \left(\mathcal{Q}^2+r^2\right)}\left(\left(\theta +r^2\right) \left(6 \pi ^2 \alpha c_2 \left(\mathcal{Q}^2 (1-2 r)+r^2\right) \left(\theta +r^2\right)+r \left(2 \pi ^2 r \left(\theta +r^2\right) \right.\right.\right.\\ \left.\left.\left.
  \left(\mathcal{Q}^2 (6 \alpha +\beta  r (r+1))+\beta  r^3\right)-3 \sqrt{\theta } M \left(\mathcal{Q}^2 (1-2 r)+r^2\right)\right)-3 M \left(\mathcal{Q}^2 (1-2 r)+r^2\right) \left(\theta +r^2\right) \cot ^{-1}\left(\frac{r}{\sqrt{\theta }}\right)\right) \right),
\end{multline}
\begin{multline}
    \label{p4}
    \omega_t(r)=\frac{1}{12 \sqrt{\theta } M r^3 \left(\mathcal{Q}^2+r^2\right)^2}\left(-6 \pi ^2 \alpha c_2 \left(\mathcal{Q}^2 r^2 (3 r-2)+\mathcal{Q}^4 (r-1) (2 r+1)-r^4\right) \left(\theta +r^2\right)^2\right.\\ \left.
    +r \left(3 \sqrt{\theta } M \left(\mathcal{Q}^2 r^2 \left((5 r-6) r^2+\theta  (3 r-2)\right)+\mathcal{Q}^4 (r-1) (\theta +r (2 \theta +r (2 r+3)))-r^4 \left(\theta +3 r^2\right)\right)\right.\right.\\ \left.\left.
    -2 \pi ^2 r \left(\theta +r^2\right)^2 \left(2 \mathcal{Q}^2 r^2 (2 \beta  r-3 \alpha )+\mathcal{Q}^4 (\beta  r (2-(r-2) r)-6 \alpha  (r-1))+2 \beta  r^5\right)\right)+3 M \left(\mathcal{Q}^2 r^2 (3 r-2)+\right.\right.\\ \left.\left.
    \mathcal{Q}^4 (r-1) (2 r+1)-r^4\right) \left(\theta +r^2\right)^2 \cot ^{-1}\left(\frac{r}{\sqrt{\theta }}\right) \right),
\end{multline}
respectively.
\begin{figure}[!htbp]
    \centering
    \includegraphics[scale=0.7]{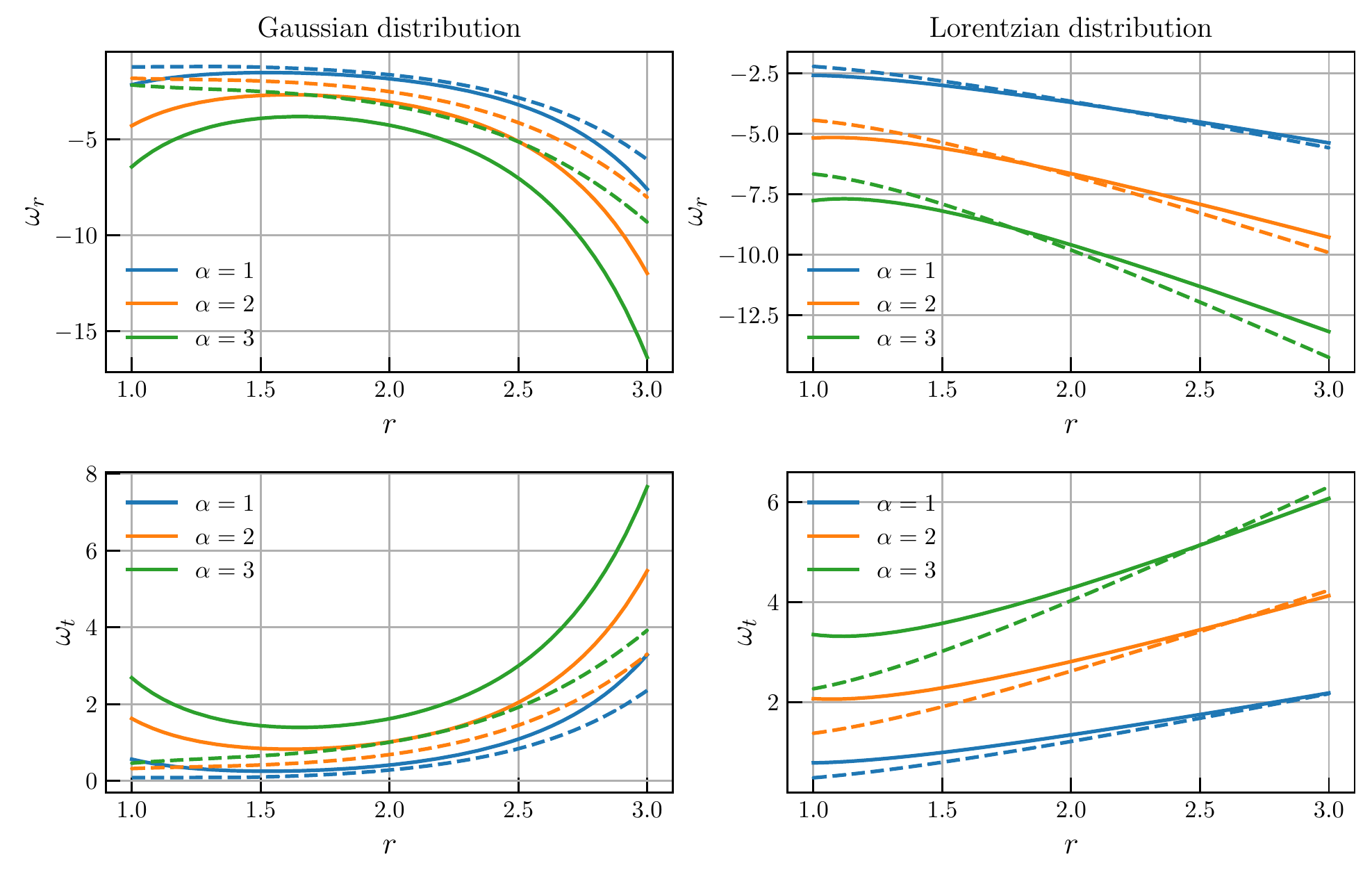}
    \caption{Linear $f(Q)$ gravity equation of state parameter for the charged traversable wormhole spacetime with both Gaussian and Lorentzian smeared mass distribution. In order to obtain results graphically, we assumed that $M=12$ and $\theta=0.5$, and also we took $\mathcal{Q}=0.1$ (solid line), $\mathcal{Q}=0.5$ (dashed line), vanishing $\beta$}
    \label{fig:999}
\end{figure}
The behaviour of EoS parameters $\omega_r(r)$ and $\omega_t(r)$ for the linear model under both distributions has been shown in Figure (\ref{fig:999}). It is observed that the radial EoS parameter decreases with $\alpha$ and radial distance increases, whereas the tangential EoS parameter increases with both radial distance and $\alpha$ increases under both Gaussian and Lorentzian distribution in this STG. Further, we study the behavior of EoS parameters for both pressures for the non-linear model under both distributions. One may check the Figure (\ref{fig:1000}), where we have shown the behavior of radial and tangential EoS parameters explicitly.

\section{Stability from TOV}\label{sec:6}

In order to discuss the equilibrium configuration for the wormhole geometry under noncommutative distributions, we shall use the generalized Tolman-Oppenheimer-Volkoff (TOV) equation of the form \cite{Saibal_2014,Kuhfittig:2020fue}
\begin{equation}\label{27}
-\frac{dp_{r}}{dr}-\frac{\Omega^{'}(r)}{2}(\rho+p_{r})+\frac{2}{r}(p_{t}-p_{r})=0,
\end{equation}
The forces namely, hydrostatic $(\mathcal{F}_H)$, the gravitational ($\mathcal{F}_G)$ and anisotropic force $(\mathcal{F}_A)$ are represented by following expressions
\begin{equation}\label{28}
F_H=-\frac{dp_{r}}{dr},\;\;\;\;\;\;\;\;F_A=\frac{2}{r}(p_{t}-p_{r}), \;\;\;\;\;\;\;\;F_G=-\frac{\Omega^{'}}{2}(\rho+p_{r}),
\end{equation}
As we are working with the constant redshift function. So, in this case the gravitational force will be vanish, i.e., $F_G=0$.\\
Thus Eq. \eqref{27} takes the form given by
\begin{equation}\label{t}
F_A+F_H=0.
\end{equation}
\begin{figure}[!htbp]
\centering
\includegraphics[scale=0.7]{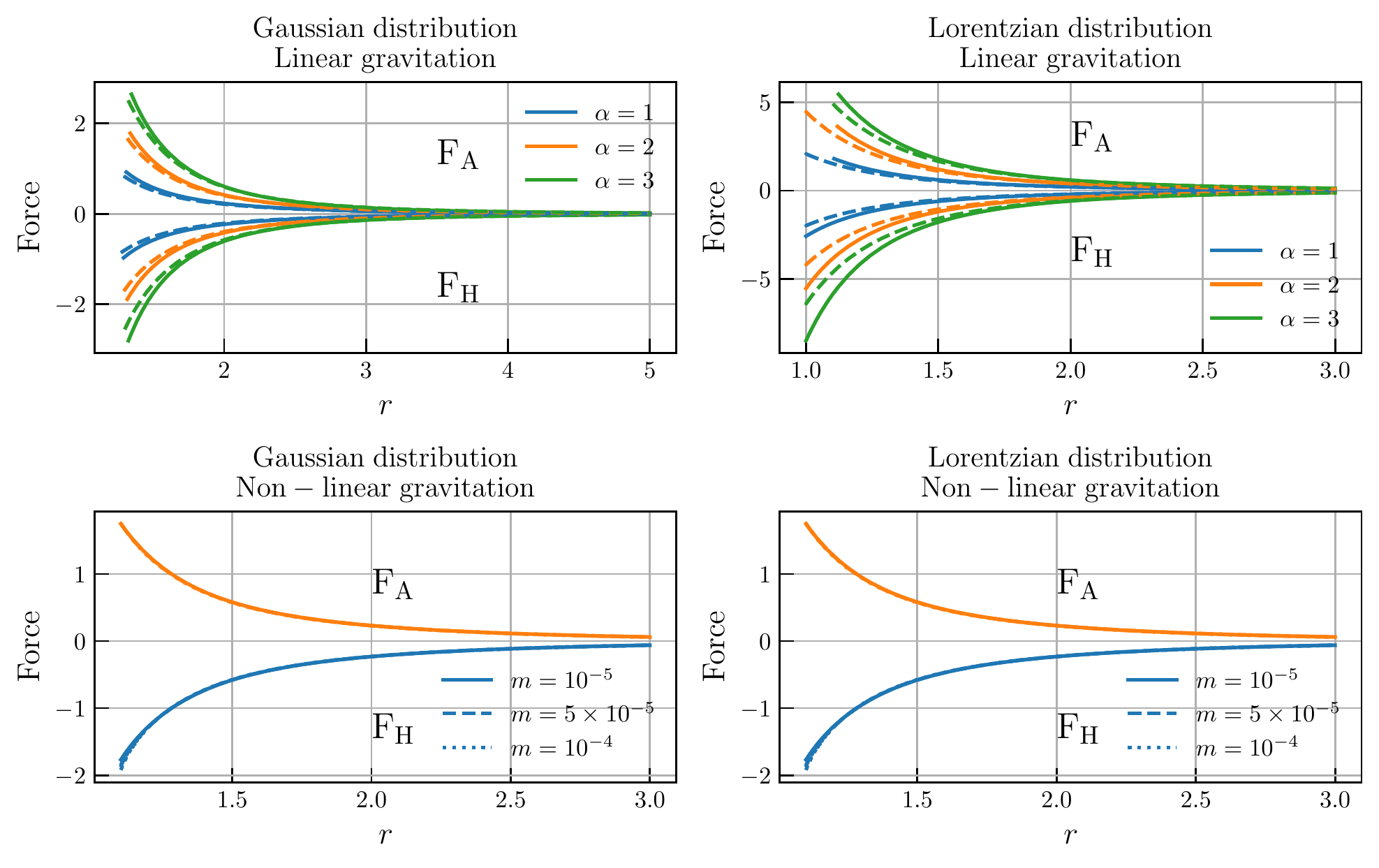}
    \caption{Behavior of TOV forces $F_H$ and $F_A$ for linear and nonlinear models under Gaussian ($left$) and Lorentzian ($right$) distributions with $M=12$, $\theta=0.5$, $n=2$, $\beta=0$ and $Q=0.1$ for non-linear case, $\mathcal{Q}=0.1$ (solid line), $\mathcal{Q}=0.5$ (dashed line) for linear case}
    \label{fig:11}
\end{figure}
In Figure \ref{fig:11}, we have depicted the behavior of the wormhole solutions for both models. For the linear model under both distributions, we found that the anisotropic force $F_A$ shows positive behavior, whereas the hydrostatic force $F_H$ shows negative behavior, i.e., both forces are identical but opposite, resulting in the equilibrium of the solutions. For non-linear cases, it is observed that both forces show the same behavior but are opposite to each other under both Gaussian and Lorentzian distributions. Hence, both models satisfy the Eq. \eqref{t} to hold the system in equilibrium. Therefore, we could conclude that our obtained wormhole solutions for both models are stable under the non-commutative framework. Readers may visit the Refs. \cite{Jawad/2016,RAHAMAN201573} where the authors have studied deeply on this topic.

\section{Concluding remarks}\label{sec:7}

A wormhole describes a shortcut distance to link different parts of the universe. To examine these solutions, the violation of NEC plays an essential role associated with the exotic matter. The usage of exotic matter would be minimized to obtain a realistic model in favor of the wormhole. This work studied the spherically symmetric static wormhole solutions in symmetric teleparallel gravity under two well-known non-commutative distributions: Gaussian and Lorentzian distributions. We have developed the field equations for the spherically symmetric wormhole spacetime metric preserving charged in $f(Q)$ gravity. Also, in this work, we have considered two WH models such as linear ($f(Q)=\alpha Q+\beta$) and non-linear ($f(Q)=Q+m Q^n$) models and studied the WH solutions under non-commutative backgrounds. It is known that the field equations of modified gravity are more complex than the field equations of GR. Although, we tried to find the exact solutions for both models. It is observed that for the linear model, we are able to find the exact solution of shape function for both distributions. But for the non-linear model, we failed to find the exact solution of shape function. For this case, we study the wormhole solutions numerically by setting some initial conditions. The graphical behaviors of our obtained solutions have been discussed below.\\
For the linear model under Gaussian distribution, we have presented the behaviors of shape functions in Figure \eqref{fig:1}. One may observe that the shape function $b(r)$ is showing increasing behavior, and the flaring out condition ($b^{'}(r_0)<1$) is also satisfied everywhere within our Lorentzian manifold. But it could be possible that for a very large value of $\mathcal{Q}$, flaring out conditions will not be satisfied on the CWH throat. Moreover, in Figure \eqref{fig:2}, we have illustrated the behavior of energy conditions (NEC, DEC, and SEC). We noticed that NEC is violated for radial pressure but satisfied for tangential pressure. It happened because of non-commutative geometry. Also, DEC is violated at each point of spacetime for radial pressure, whereas tangential DEC could be satisfied for relatively small and positive values of $\alpha$ under the Gaussian framework. Again for the non-linear model under Gaussian distribution, in Figures \eqref{fig:333} and \eqref{fig:444}, we have depicted the behavior of shape functions and energy conditions. We have noticed during the numerical analysis that the flaring out condition is satisfied at the CWH throat for small values of charge $\mathcal{Q}$ on the asymptotically flat CWH region. Also, for any positive values of charge $\mathcal{Q}$, NEC is violated for the radial pressure and satisfied for the tangential one. SEC was also violated in the vicinity of WH's throat. Violation of NEC confirms the presence of exotic matter at the WH throat, which is necessary for the traversability of WH.\\
Moving forward, for the linear model under the Lorentzian source, we have plotted the graphs for the shape functions and energy conditions presented in Figures \eqref{fig:3} and \eqref{fig:4}. It is obvious that the flaring out condition is satisfied throughout the spacetime. But for large $\alpha$ and $\mathcal{Q}$, flaring out the condition will not be validated anymore. Also, we noticed that NEC is violated for radial pressure and obeyed for tangential pressure for any charge $\mathcal{Q}\geq0$ as well as for any positive $\alpha$. DEC is violated for radial pressure in the entire spacetime but validated for tangential pressure only for $\alpha\leq 1$. SEC is also violated even further from the charged WH throat because of the smeared mass. Furthermore, for the non-linear model under Lorentzian distribution, we found that the flaring out condition is violated near the WH troat but validated at the throat for any $\mathcal{Q}\geq 0$. Moreover, we observed that for the non-linear $f(Q)$ model, all the energy conditions were violated for any $\mathcal{Q}\geq 0$. One may check the Figures (\ref{fig:777}) and (\ref{fig:888}) for more details.\\
Further, we have studied the behavior of EoS parameters $\omega$ for radial and tangential pressures with different charges $\mathcal{Q}$ for both models. The graphical overview of the behavior of the EoS parameter for both pressures under both distributions have been shown in Figures (\ref{fig:999}) and (\ref{fig:1000}). Lastly, we have used a tool called the TOV equation to check the stability of our obtained WH solutions for both models. From our obtained graph (see Figure \ref{fig:11}), we found that both models are stable under both distributions.\\
In addition, it will be interesting to compare the results of charged wormholes with non-charged case within the same choices of $f(Q)$ form. Morris-Thorne (MT) wormholes within the linear $f(Q)=\alpha Q$ model were successfully probed in the paper \cite{Zinnat/2021}. In relation to the charged wormhole solutions, presented in the current work, tangential NEC were validated, as well as DEC, which greatly coincides with our data (however, in our case tangential DEC is validated only for relatively small and positive values of $\alpha$ assuming vanishing $\beta$). In the aforementioned paper, quadratic gravity were also investigated in details. For non-linear case with special form of shape function $b(r)$, radial NEC were violated everywhere and tangential NEC were validated for some forms of shape function $b(r)$, which could be also related to our case. But, as it was revealed, DEC for non-linear gravity Morris-Thorne wormhole does deviate from our data (for MT WH radial DEC was validated instead of tangential one).\\
Thus, it would be interesting to mention that our obtained results are consistent with the non-commutative framework in the symmetric teleparallel gravity. Also, It would be more interesting to explore wormhole solutions in this modified $f(Q)$ gravity by taking other matter sources into account.

\section*{Data Availability Statement}
There are no new data associated with this article.

\section*{Acknowledgments}
PKS acknowledges National Board for Higher Mathematics (NBHM) under Department of Atomic Energy (DAE), Govt. of India for financial support to carry out the Research project No.: 02011/3/2022 NBHM(R.P.)/R\&D II/2152 Dt.14.02.2022. We are very much grateful to the honorable referee and to the editor for the illuminating suggestions that have significantly improved our work in terms
of research quality, and presentation. 
\end{document}